\documentstyle[prb,aps,epsf]{revtex}

\begin{document}
\draft

\advance\textheight by 0.5in \advance\topmargin by -0.in

\twocolumn[\hsize\textwidth\columnwidth\hsize\csname@twocolumnfalse\endcsname{ }

\title{Universality of electron correlations in conducting carbon nanotubes}
\author{Arkadi A. Odintsov$^{1,2}$ and Hideo Yoshioka$^{2,3}$}
\address{
$^{1}$NEC Research Institute, 4 Independence Way, Princeton, New Jersey 08540
\\
$^{2}$Department of Applied Physics, Delft University of Technology, 
2628 CJ Delft, The Netherlands. 
\\
$^{3}$Department of Physics, Nagoya University,
Nagoya 464-8602, Japan. }
\date{\today}
\maketitle

\begin{abstract}
Effective low-energy Hamiltonian 
of interacting electrons in conducting single-wall
carbon nanotubes with arbitrary chirality is derived from the microscopic
lattice model. The parameters of the Hamiltonian show very weak dependence
on the chiral angle, which makes the low energy properties of conducting
chiral nanotubes universal. The strongest Mott-like electron instability
at half filling is investigated within the self-consistent harmonic 
approximation. The energy gaps occur in all modes of elementary excitations 
and estimate at $0.01-0.1$ eV.
\end{abstract}

\vskip -0.5 truein
]

\smallskip Single wall carbon nanotubes (SWNTs) are linear macromolecules
whose individual properties can be studied by methods of nanophysics \cite
{Thess}. Recent demonstration of electron transport through single \cite
{Tans} and multiple \cite{Bockrath} SWNTs has been followed by remarkable
observations of atomic structure \cite{Wildoer,Johnson}, one-dimensional van
Hove singularities \cite{Wildoer}, standing electron waves \cite{Venema}
and, possibly, electron correlations \cite{Tans2} in these systems.
Moreover, the first prototype of a functional device - the nanotube field
effect transistor working at room temperature - has been fabricated recently 
\cite{Tans3}.

Structurally uniform SWNTs can be characterized by the wrapping vector 
$\vec{w}=N_{1}\vec{a}_{1}+N_{2}\vec{a}_{2}$ 
given by the linear combination of
primitive lattice vectors $\vec{a}_{\pm}=(\pm 1,\sqrt{3})a/2$,
with $a \approx 0.246$ nm (Fig. 1). 
It is natural to separate
non-chiral armchair ($N_{1}=N_{2}$) and zig-zag ($N_{1}=-N_{2}$) nanotubes
from their chiral counterparts. Recent scanning tunneling microscopy study 
\cite{Wildoer} has revealed that individual SWNTs are generally chiral.
According to the single-particle model, the nanotubes with $N_{1}-N_{2}=0$
mod $3$ have gapless energy spectrum \cite{commgapless} and are therefore
conducting; otherwise, the energy spectrum is gapped and SWNTs are
insulating. Therefore, on the level of non-interacting electrons, physical
properties of SWNTs are determined by their geometry.

The Coulomb interaction in one-dimensional SWNTs should result in a variety
of correlation effects due to the non-Fermi liquid ground state of the
system. In particular, metallic armchair SWNTs are predicted to be Mott
insulating at half-filling \cite{Balents-Fisher},\cite{Krotov-Lee-Louie},%
\cite{Kane-Balents-Fisher},\cite{Yoshioka}.
Upon doping the nanotubes become
conducting but still display 
density wave instabilities 
in three modes of elementary excitations 
with neutral total charge
\cite{Egger-Gogolin},\cite{Yoshioka}.

Experimental observation of electron correlations still remains challenging
since their signatures are usually masked by charging effects. Two recent
transport spectroscopy experiments on an individual conducting SWNT \cite
{Tans2} and a rope of SWNTs \cite{Cobden} produced contradicting results.
The data by Tans et. al. \cite{Tans2} assumes spin polarized tunneling into
a nanotube, which in turn suggests the interpretation in terms of electron
correlations. On the other hand, the data by Cobden et. al. \cite{Cobden}
fits the constant interaction model remarkably well and shows no
signatures of exotic correlation effects.
Since the  atomic structure of particular 
SWNTs studied in these experiments is not known,
it might be appealing to interpret the discrepancy in the
results in terms of
geometry-dependent {\it many-particle} properties of
conducting SWNTs.

Unfortunately, consistent theory of interacting electrons in
chiral SWNTs is lacking, despite such a theory has been recently developed
for armchair nanotubes \cite{Egger-Gogolin},\cite{Kane-Balents-Fisher},\cite
{Yoshioka},\cite{comment1}. In this work we establish effective low-energy
model for conducting chiral SWNTs and evaluate its parameters (scattering
amplitudes) from the microscopic theory. We found very weak dependence of
the dominant scattering amplitudes on the chiral angle. 
This allows us to introduce  {\it universal} low-energy Hamiltonian  
of conducting SWNTs.
According to the results of the renormalization group analysis
\cite{Yoshioka} the strongest Mott-like electron instability occurs at
half-filling. We investigate this instability 
using the self-consistent harmonic approximation.
Substantial energy gaps are found 
in all  modes of elementary excitations.
The conditions for experimental observation of the gaps
are briefly discussed.

We start from the standard kinetic term $H_{k}$ of the
tight-binding Hamiltonian for $p_{z}$ electrons 
of a graphite sheet \cite{Wallace}, 
\begin{equation}
H_{k}=\sum_{s,\vec{k}}\left\{ \xi (\vec{k})a_{-,s}^{\dagger }(\vec{k}%
)a_{+,s}(\vec{k})+h.c.\right\} .  \label{eqn:1}
\end{equation}
Here $a_{p,s}(\vec{k})$ are the Fermi operators for electrons at the
sublattice $p=\pm $ (Fig. 1) with the spin $s=\pm $ and the wavevector $\vec{%
k}=(k_{x},k_{y})$. The matrix elements are given by $\xi (\vec{k})=-t({\rm e}%
^{-{\rm i}k_{y}a/\sqrt{3}}+2{\rm e}^{{\rm i}k_{y}a/2\sqrt{3}}\cos k_{x}a/2)$%
, $t$ being the hopping amplitude between neighboring atoms. The eigenvalues
of the Hamiltonian vanish at two Fermi points of the Brillouin zone, 
$\alpha \vec{K}$ with $\alpha =\pm $ and 
$\vec{K}=(K,0)$, $K=4\pi /3a$.

We consider conducting chiral $(N_{1},N_{2})$ SWNT of the radius 
$R=(a/2\pi)\sqrt{N_{1}^{2}+N_{1}N_{2}+N_{2}^{2}}$ whose 
axis  $x^{\prime }$ forms the
angle $\chi =\arctan [(N_{2}-N_{1})/\sqrt{3}(N_{1}+N_{2})]$ with the
direction of chains of carbon atoms ($x$ axis in Fig. 1). Expanding Eq. (%
\ref{eqn:1}) near the Fermi points to the lowest order in $\vec{q}=\vec{k}%
-\alpha \vec{K}=q(\cos \chi ,\sin \chi )$ and introducing slowly varying
Fermi fields $\psi _{p\alpha s}(x^{\prime })=L^{-1/2}\sum_{q=2\pi
n/L}e^{iqx^{\prime }}a_{p\alpha s}(\vec{q}+\alpha \vec{K})$, we obtain, 
\begin{figure}[tb]
\epsfxsize=\columnwidth\epsfbox{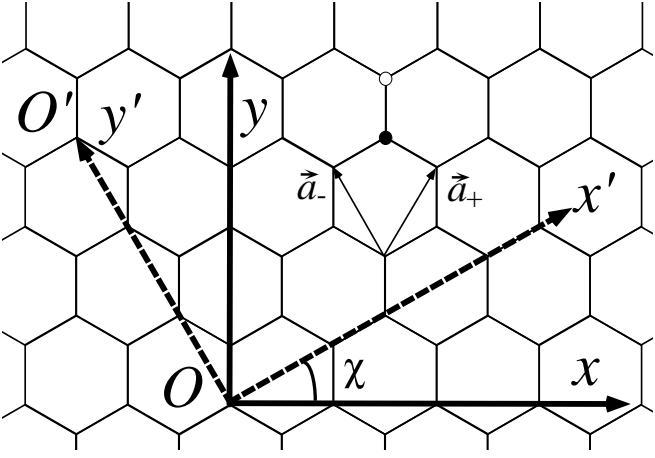}
\vspace{15pt}
\caption{Graphite lattice consists of two atomic sublattices $p=+,-$ denoted
by filled and open circles. SWNT at the angle $\chi $ to $x$ axis can be
formed by wrapping the graphite sheet along $\vec{w}=OO'$ vector. }
\label{fig1}
\end{figure}
\noindent
\begin{equation}
H_{k}=-iv\sum_{p\alpha s}\alpha e^{-ip\alpha \chi }\int {\rm d}%
x^{\prime }\psi _{p\alpha s}^{\dagger }\partial _{x^{\prime }}\psi
_{-p\alpha s},  \label{Hk}
\end{equation}
with the Fermi velocity $v=\sqrt{3}ta/2\approx 8.1\times 10^{5}$ m/s. The
kinetic term can be diagonalized by the unitary transformation 
\begin{equation}
\psi _{p\alpha s}=\frac{1}{\sqrt{2}}e^{-ip\alpha \chi /2}\sum_{r=\pm
}(r\alpha )^{\frac{1-p}{2}}\varphi _{r\alpha s}  \label{Utrans}
\end{equation}
to the basis $\varphi _{r\alpha s}$ of left- ($r=-$) and right- ($r=+$)
movers.

\strut The Coulomb interaction has the form 
\begin{eqnarray}
H_{int} &=&\frac{1}{2}\sum_{pp^{\prime },\left\{ \alpha _{i}\right\}
,ss^{\prime }}V_{pp^{\prime }}(2\bar{\alpha}K)  \nonumber  \label{Hint} \\
&&\times \int dx^{\prime }\psi _{p\alpha _{1}s}^{\dagger }\psi _{p^{\prime
}\alpha _{2}s^{\prime }}^{\dagger }\psi _{p^{\prime }\alpha _{3}s^{\prime
}}\psi _{p\alpha _{4}s},
\label{Hint}
\end{eqnarray}
with the matrix elements 
\begin{equation}
V_{pp^{\prime }}(2\bar{\alpha}K)=\frac{1}{\rho }\sum_{\vec{r}_{p}}U(\vec{r}%
_{p}-\vec{r}_{p^{\prime }})\exp [-2i\bar{\alpha}\vec{K}(\vec{r}_{p}-\vec{r}%
_{p^{\prime }})],  \label{Vpp'}
\end{equation}
corresponding to the amplitudes of intra- ($p=p^{\prime }$) and inter- ($%
p=-p^{\prime }$) sublattice forward ($\bar{\alpha}=0$) and backward ($\bar{%
\alpha}=\pm 1$) scattering \cite{commfbs} (the sum is taken over the nodes of
the sublattice $p$ of SWNT). Here $U(\vec{r})={e^{2}}/\{\kappa \sqrt{%
a_{0}^{2}+(x^{\prime })^{2}+4R^{2}\sin ^{2}(y^{\prime }/2R)}\}$ is the
Coulomb interaction with a short-distance cutoff $%
a_{0}\sim a$, 
$\rho =4\pi R/\sqrt{3}a^{2}$ is a linear density of
sublattice nodes along SWNT, and $\bar{\alpha}=(\alpha _{1}-\alpha
_{4})/2=(\alpha _{3}-\alpha _{2})/2$. We will choose the parameter $a_{0}$
from the requirement that the on-site interaction in the original
tight-binding model corresponds to the difference between the ionization
potential and electron affinity of $sp^{2}$ hybridized carbon \cite{Moore}.
This procedure gives $a_{0}=0.526a$.

The forward scattering part $H_{F}$ 
[terms with $\bar{\alpha}= 0$ in Eq. (\ref{Hint})]
of the Hamiltonian $H_{int}$
can be separated into the Luttinger
model-like term $H_{\rho }$ and the term $H_{f}$ related to the difference $%
\Delta V(0)={V}_{pp}(0)-{V}_{p-p}(0)$ between intra- and intersublattice
amplitudes,
\begin{eqnarray}
H_{\rho } &=&\frac{V_{pp}(0)}{2}\int {\rm d}x^{\prime}\rho ^{2}(x^{\prime}), \\
H_{f} &=&-\frac{\Delta V(0)}{2}\sum_{p\alpha \alpha ^{\prime
}ss^{\prime }}\int {\rm d}x^{\prime}
\psi _{p\alpha s}^{\dagger }\psi _{-p\alpha
^{\prime }s^{\prime }}^{\dagger }\psi _{-p\alpha ^{\prime }s^{\prime }}\psi
_{p\alpha s},
\end{eqnarray}
where $\rho =\sum_{p\alpha s}\psi _{p\alpha s}^{\dagger }\psi _{p\alpha s}$
is the total electron density. The backscattering Hamiltonian $H_{B}$ [terms
with $\bar{\alpha}= \pm 1$
in Eq. (\ref{Hint})] can be subdivided into the intrasublattice $H%
_{b}^{(+)}$ ($p=p^{\prime }$) and intersublattice $H_{b}^{(-)}$ ($%
p=-p^{\prime }$) parts.

\strut The dominant contribution to the forward scattering amplitudes $%
V_{pp^{\prime }}(0)$ comes from the long range component of the Coulomb
interaction, $V_{pp}(0)=(2e^{2}/\kappa) \ln (R_{s}/R)$, where $R_{s}\simeq
\min (L,D)$ characterizes the screening of the interaction due to a finite
length $L$ of the SWNT and/or the presence of metallic electrodes at a
distance $D$ \cite{Kane-Balents-Fisher}. The forward scattering differential
part $\Delta V(0)$ and the intrasublattice backscattering amplitude $%
V_{pp}(2K)$ can be estimated from Eq. (\ref{Vpp'}) as follows,
$\Delta V(0),V_{pp}(2K) $ $\sim ae^{2}/\kappa R$ (for $a_{0}\sim a$). Despite the
amplitudes $\Delta V(0),V_{pp}(2K)$ are much smaller than $V_{pp}(0),$ they
produce essentially non-Luttinger terms in the low-energy Hamiltonian
which will be important in further analysis.

We evaluated the matrix elements (\ref{Vpp'})\ numerically for chiral SWNTs
with radiuses $R$ in the range $2R/a=4-7$ ($2R/a=5.5$ for (10,10)
SWNTs). We found that dimensionless amplitudes $2 \pi \kappa R[\Delta
V(0),V_{pp}(2K)]/ae^{2}$ show very weak dependence on the radius of SWNT and
its chiral angle (see Table 1). The results are sensitive to the value of
the cutoff parameter $a_{0}$.

The intersublattice backscattering amplitude $V_{p-p}(2K)$ is almost three
orders of magnitude smaller than $\Delta V(0)$, $V_{pp}(2K)$. This is due to
the $C_{3}$ symmetry of a graphite lattice, which leads to an exact
cancellation of the terms (\ref{Vpp'}) contributing to $V_{p-p}(2K)$ in the
case of a plane graphite sheet. 
The matrix elements $V_{p-p}(2K)$ are generally
complex due to asymmetry of effective 1D intersublattice interaction
potential (the matrix elements are real for symmetric zig-zag and armchair
SWNTs). Let us note that after the unitary transformation (\ref{Utrans}) of
the Hamiltonian $H=H_{k}+H_{int}$, 
the chiral angle $\chi $ enters {\it only} to the intersublattice
backscattering matrix elements \cite{comminvar}. Due to the smallness of
these matrix elements, the low-energy properties of chiral SWNTs are
expected to be virtually independent of the chiral angle.
\begin{table}[bt] \centering%
\begin{tabular}{|l|c|c|c|}
$a_{0}/a$ & $\Delta V(0)$ & $V_{pp}(2K)$ & $|V_{p-p}(2K)|$ \\ \hline
$0.4$ & $0.44265-0.44274$ & $0.97060-0.97095$ & $0.6-2.2\times
10^{-3}$ \\ 
$0.526$ & $0.17378-0.17395$ & $0.53549-0.53561$ & $0.5-1.6\times
10^{-3}$ \\ 
$0.7$ & $0.04880-0.04895$ & $0.24778-0.24797$ & $0.3-1.5\times
10^{-3}$ \\ 
\end{tabular}
\vspace{0.5cm}
\caption{Scattering amplitudes $\Delta V(0)$, $V_{pp}(2K)$, $V_{p-p}(2K)$ 
in units $ae^{2}/2\pi\kappa R$ for all chiral SWNTs 
with radiuses $R$ in the range $2R/a=4-7$.
\label{tab1}}%
\end{table}%

Neglecting the intersublattice backscattering we arrive to the universal
low-energy model of conducting SWNTs given by the Hamiltonian $%
H=H_{k}+H_{F}+H_{b}^{(+)}$. The latter can be bosonized along the lines of
Refs. \cite{Egger-Gogolin,Yoshioka}. We introduce bosonic representation of
the Fermi fields, 
\begin{equation}
\varphi _{r\alpha s}=\frac{\eta _{r\alpha s}}{\sqrt{2\pi \tilde{a}}}\exp
\left[ {\rm i}rq_{F}x^{\prime}
+\frac{{\rm i}r}{2}\left\{ \theta _{\alpha s}+r\phi
_{\alpha s}\right\} \right] ,
\end{equation}
and decompose the phase variables $\theta _{\alpha s},\phi _{\alpha s}$ into
symmetric $\delta =+$ and antisymmetric $\delta =-$ modes of the charge $%
\rho $ and spin $\sigma $ excitations, $\theta _{\alpha s}=\theta _{\rho
+}+s\theta _{\sigma +}+\alpha \theta _{\rho -}+\alpha s\theta _{\sigma -}$
and $\phi _{\alpha s}=\phi _{\rho +}+s\phi _{\sigma +}+\alpha \phi _{\rho
-}+\alpha s\phi _{\sigma -}$. The bosonic fields satisfy the commutation
relation, 
$[\theta _{j\delta }(x_{1}),\phi _{j^{\prime }\delta ^{\prime
}}(x_{2})]={\rm i}(\pi /2){\rm sign}(x_{1}-x_{2})\delta _{jj^{\prime
}}\delta _{\delta \delta ^{\prime }}$. 
The Majorana fermions $\eta _{r\alpha
s}$ are introduced \cite{Egger-Gogolin} 
to ensure correct anticommutation rules for different
species $r,\alpha ,s$ of electrons, and satisfy $[\eta _{r\alpha s},\eta
_{r^{\prime }\alpha ^{\prime }s^{\prime }}]_{+}=2\delta _{rr^{\prime
}}\delta _{\alpha \alpha ^{\prime }}\delta _{ss^{\prime }}$.  
The quantity $q_{F}=\pi n/4$ is related to the
deviation $n$ of the average electron density from half-filling, 
and $\tilde{a}\sim a$ is the standard ultraviolet cutoff.

The universal low-energy Hamiltonian of conducting SWNTs has the following
bosonized form, 
\begin{eqnarray}
H &=&\sum_{j=\rho ,\sigma }\sum_{\delta =\pm }\frac{v_{j\delta }}{%
2\pi }\int {\rm d}x^{\prime}
\left\{ K_{j\delta }^{-1}(\partial _{x^{\prime}}\theta _{j\delta
})^{2}+K_{j\delta }(\partial _{x^{\prime}}\phi _{j\delta })^{2}\right\}  
\nonumber \\
&&+\frac{1}{2(\pi \tilde{a})^{2}}
\int {\rm d}x^{\prime}\{[\Delta V(0)-V_{pp}(2K)] 
\nonumber \\
&&[\cos (4q_{F}x^{\prime}+2\theta _{\rho +})\cos 2\theta _{\sigma +}
-\cos 2\theta_{\rho -}\cos 2\theta _{\sigma -}]  
\nonumber \\
&&-\Delta V(0)\cos (4q_{F}x^{\prime}+2\theta _{\rho +})\cos 2\theta _{\rho -} 
\nonumber \\
&&+\Delta V(0)\cos (4q_{F}x^{\prime}+2\theta _{\rho +})\cos 2\theta _{\sigma -} 
\nonumber \\
&&-{\Delta V(0)}\cos 2\theta _{\sigma +}\cos 2\theta _{\rho -}  
\nonumber \\
&&+{\Delta V(0)}\cos 2\theta _{\sigma +}\cos 2\theta _{\sigma -}  
\nonumber \\
&&-{V}_{pp}{(2K)}\cos (4q_{F}x^{\prime}+2\theta _{\rho +})\cos 2\phi _{\sigma -} 
\nonumber \\
&&+{V}_{pp}{(2K)}\cos 2\theta _{\sigma +}\cos 2\phi _{\sigma -}  \nonumber \\
&&+{V}_{pp}{(2K)}\cos 2\theta _{\rho -}\cos 2\phi _{\sigma -}  \nonumber \\
&&+{V}_{pp}{(2K)}\cos 2\theta _{\sigma -}\cos 2\phi _{\sigma -}\},
\label{Hbos}
\end{eqnarray}
$v_{j\delta }=v\sqrt{A_{j\delta }B_{j\delta }}$ and $K_{j\delta }=\sqrt{%
B_{j\delta }/A_{j\delta }}$ being the velocities of excitations and
exponents for the modes $j,\delta $. The parameters $A_{j\delta }$, $%
B_{j\delta }$ are given by%
\begin{eqnarray}
A_{\rho +} &=&1+\frac{4\bar{V}(0)}{\pi v}-\frac{\Delta V(0)}{4\pi v}-\frac{%
V_{pp}(2K)}{2\pi v},  \label{Arp} \\
A_{\nu \delta } &=&1-\frac{\Delta V(0)}{4\pi v}-\delta \frac{V_{pp}(2K)}{%
2\pi v},  \label{And} \\
B_{\nu \delta } &=&1+\frac{\Delta V(0)}{4\pi v}.  \label{Bnd}
\end{eqnarray}
with $\bar{V}(0)=[V_{pp}(0)+V_{p-p}(0)]/2$.

Since $\bar{V}(0)/v\sim \ln (R_{s}/R)$ $\gg 1$ and 
$\Delta V(0)/v$, $V_{pp}(2K)/v\sim a/R$ $\ll 1$, 
the renormalization of the parameters $K_{j\delta }$, $v_{j\delta }$
by the Coulomb interaction is the strongest in $\rho +$ mode. Assuming $%
\kappa =1.4$ \cite{Egger-Gogolin} and $R_{s}=100$ nm we obtain $K_{\rho
+}\simeq 0.2, v_{\rho +}\simeq v/K_{\rho +}$ for generic SWNTs with $R\sim 0.7$
nm \cite{Kane-Balents-Fisher}. The interaction in the other modes is weak: $%
v_{j\delta }=v,$ $K_{j\delta }=1,$ up to a factor $1+O(a/R).$

The renormalization group analysis 
of armchair SWNTs with long range Coulomb interaction has
been performed in Refs. \cite{Yoshioka},\cite
{Kane-Balents-Fisher}. 
The modification of the parameters of the Hamiltonian 
(\ref{Hbos}) by the neglected small term $V_{p-p}(2K)$ 
should not change the results qualitatively \cite{commRG}.
The most relevant perturbation is the umklapp scattering at
half-filling. In this case
the non-linear terms of the Hamiltonian (\ref{Hbos}), which do not contain $%
\cos 2\theta _{\sigma -}$ scale to strong coupling and the phases $\theta
_{\rho +},\theta _{\sigma +},\theta _{\rho -},\phi _{\sigma -}$ get locked
at $(\theta _{\rho +}^{(m)},\theta _{\sigma +}^{(m)},\theta _{\rho
-}^{(m)},\phi _{\sigma -}^{(m)})=(0,0,0,0)$ or $(\pi /2,\pi /2,\pi /2,\pi
/2) $. Therefore, the ground state of half-filled SWNT is the Mott insulator
with all kinds of the excitations gapped. 

To estimate the gaps
quantitatively, we will employ 
the self-consistent harmonic approximation
which follows from Feynman's variational principle \cite{Feynman}.
We consider trial harmonic Hamiltonian of the form:
\begin{eqnarray}
H_{0} 
= \sum_{j\delta}\frac{v_{j\delta }}{2\pi }
\int {\rm d}x^{\prime}
&\{ & 
	K_{j\delta }^{-1}
	[(\partial _{x^{\prime}}\theta _{j\delta})^{2}
	+(1-\delta_{j\sigma}\delta_{\delta -})
	q_{j\delta}^{2}\theta _{j\delta }^{2}] 
\nonumber \\
&	+ &
	K_{j\delta }
	[(\partial_{x^{\prime}}\phi _{j\delta })^{2}
	+ \delta_{j\sigma} \delta_{\delta -}
	q_{j\delta}^{2}\phi _{j\delta }^{2}]
\},  
\label{Hscha}
\end{eqnarray}
$q_{j\delta }$ being variational parameters. By minimizing the upper
estimate for the Free energy $F^{*}=F_{0}+\left\langle H-H_{0}\right\rangle
_{0}$ \cite{Feynman} one obtains the following self-consistent
equations, 
\begin{eqnarray}
q_{\rho +}^{2} &=&\frac{2K_{\rho +}}{\pi \tilde{a}^{2}v_{\rho +}}c_{\rho
+}\left\{ [V_{pp}(2K)-\Delta V(0)]c_{\sigma +}\right.  
\nonumber \\
&&\left. +\Delta V(0)c_{\rho -}+V_{pp}(2K)d_{\sigma -}\right\} , 
\label{qrp} \\
q_{\rho -}^{2} &=&\frac{2K_{\rho -}}{\pi \tilde{a}^{2}v_{\rho -}}c_{\rho
-}\left\{ \Delta V(0)c_{\rho +}+\Delta V(0)c_{\sigma +}\right.  
\nonumber \\
&&\left. -V_{pp}(2K)d_{\sigma -}\right\} , 
\label{qrm} \\
q_{\sigma +}^{2} &=&\frac{2K_{\sigma +}}{\pi \tilde{a}^{2}v_{\sigma +}}%
c_{\sigma +}\left\{ [V_{pp}(2K)-\Delta V(0)]c_{\rho +}\right.  
\nonumber \\
&&\left. +\Delta V(0)c_{\rho -}-V_{pp}(2K)d_{\sigma -}\right\} , 
\label{gsp} \\
q_{\sigma -}^{2} &=&\frac{2}{\pi \tilde{a}^{2}K_{\sigma -}v_{\sigma -}}%
d_{\sigma -}V_{pp}(2K)\left\{ c_{\rho +}-c_{\sigma +}-c_{\rho -}\right\} ,
\label{qsm}
\end{eqnarray}
where $c_{j\delta }=\left\langle \cos 2\theta _{j\delta
}\right\rangle _{0}=\cos 2\theta _{j\delta }^{(m)}(\gamma \tilde{a}%
q_{j\delta })^{K_{j\delta }}$, $d_{\sigma -}=\left\langle \cos 2\phi
_{\sigma -}\right\rangle _{0}=\cos 2\phi _{\sigma -}^{(m)}(\gamma \tilde{a}%
q_{\sigma -})^{1/K_{\sigma -}}$, $\left\langle ...\right\rangle _{0}$
denotes averaging with respect to the trial Hamiltonian (\ref{Hscha}), and $%
\gamma \simeq 0.890$ for the exponential ultraviolet cutoff. 
Note that $%
\left\langle \cos 2\theta _{\sigma -}\right\rangle _{0}=0$, so that only the
terms of the Hamiltonian (\ref{Hbos}) which scale to the strong coupling
contribute to Eqs. (\ref{qrp})-(\ref{qsm}).

In the limiting case of interest, $|\Delta V(0)|, |V_{pp}(2K)|\ll v$ and $%
K_{\rho +}\ll 1$, the solution of 
Eqs. (\ref{qrp})-(\ref{qsm}) 
can be found in a closed form, giving rise to the following
estimates for the gaps $\Delta _{j\delta }=v_{j\delta }q_{j\delta }$ in the
energy spectra,
\begin{eqnarray}
\Delta _{\rho +} &=&\frac{v_{\rho +}}{\gamma \tilde{a}}\left( \frac{2\gamma
^{2}V_{\rho +}}{\pi v_{\rho +}}\right) ^{1/(1-K_{\rho +})}  \label{drp}
\\
\Delta _{\rho -} &=&\frac{|\Delta V(0)|}{V_{\rho +}}\Delta _{\rho +}
\label{drm} \\
\Delta _{\sigma +} &=&\frac{|V_{pp}(2K_{0})-\Delta V(0)|}{V_{\rho +}}\Delta
_{\rho +}  \label{dsp} \\
\Delta _{\sigma -} &=&\frac{|V_{pp}(2K_{0})|}{V_{\rho +}}\Delta _{\rho +}
\label{dsm}
\end{eqnarray}
\begin{figure}[tb]
\epsfxsize=\columnwidth\epsfbox{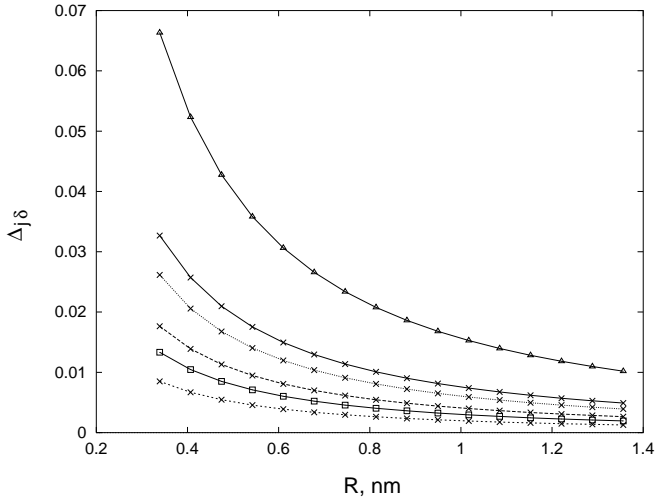}
\vspace{25pt}
\caption{The energy gaps $\Delta _{j\delta }$ for the modes $\rho +$, 
$\sigma -$, $\sigma +$, $\rho -$ at $a_{0}=0.526a$ (lines marked by
crosses, from top to bottom) and for the mode $\rho +$ at $a_{0}=0.4a$
(triangles) and at $a_{0}=0.7a$ (squares). The energy is in units $\hbar v/%
\tilde{a}\simeq 2.16$ eV for $\tilde{a}=a$.}
\label{fig2}
\end{figure}
\noindent
with 
$V_{\rho +}=\left\{ [\Delta V(0)-V_{pp}(2K)]^{2}+[\Delta V(0)]^{2}\right.$ 
$\left.+[V_{pp}(2K)]^{2}\right\} ^{1/2}$. 
In the above expressions we used the
approximation, $v_{0}/v_{\rho +}=K_{\rho +}$ and $v_{0}/v_{j\delta
}=K_{j\delta }=1$ for $\sigma \pm $ and $\rho -$ modes. The formulae (\ref
{drp})-(\ref{dsm}) indicate that the largest gap occurs in $\rho +$ mode,
albeit all four gaps are of the same order for realistic values of the
matrix elements (see Table 1). The gaps decrease as $\Delta _{j\delta
}\propto (1/R)^{1/(1-K_{\rho +})}\simeq (1/R)^{5/4}$ with the tube radius.
This should be contrasted to the $1/R$ dependence of wide semiconductor
gaps and $1/R^{2}$ dependence of narrow deformation induced gaps 
\cite{Kane-Mele} expected
from the single particle picture.

In Figure 2 we present numerical results for the gaps $\Delta _{j\delta }$
for the cutoff parameter $a_{0}=0.526a$. 
The data for somewhat larger and somewhat smaller values of $a_{0}$
indicate possible variation of the gap $\Delta _{\rho +}$ due to
uncertainty in the short distance cutoff of the Coulomb interaction.
The gaps can be loosely estimated at 
$\Delta _{j\delta}\sim 0.01-0.1$ eV
for typical SWNTs with $R \simeq 0.7$ nm.
Due to the gaps in the spectrum of bosonic elementary excitations,
the  electronic density of states should 
disappear in the subgap region and display features
at the gap frequencies and their harmonics. 
Both signatures should be observable by means 
of the tunneling spectroscopy.

Why have the gaps not been observed in the experiments
\cite{Tans2,Cobden,Wildoer}?
This might be due to the effect of  metallic electrodes.
The difference in the workfunctions
of the electrodes (Au, Pt) and the nanotube
results in a downward shift of
the Fermi level of the nanotube by
a few tenths of an eV  \cite{Wildoer}.
This causes substantial  deviation 
$\Delta n = 4q_{F}/\pi \sim  1$ nm$^{-1}$
of the electron density in SWNT from half 
filling, at least in the vicinity of the electrodes.
Therefore, we expect the gap features to be observable in
the layouts with well separated
(to a distance $d \gg \hbar v_{F} / \Delta_{\rho +}$)
source and drain contacts.
The piece of nanotube between them
should be well isolated from any conductor.

In conclusion, we have developed effective low-energy theory of 
conducting chiral SWNTs with the long-range Coulomb interaction. 
The many-particle properties of SWNTs 
are found to be virtually independent of the chiral angle.
The universal Hamiltonian (\ref{Hbos}) of conducting SWNTs is introduced.
The Mott-like energy gaps in the range of
$0.01-0.1$ eV should be observable at half filling.   

The authors would like to thank B.L. Altshuler, G.E.W. Bauer, R. Egger,
Yu.V. Nazarov, and N. Wingreen for stimulating discussions. The
financial support of the Royal Dutch Academy of Sciences
(KNAW) is gratefully acknowledged. 
One of us (A.O.) acknowledges the kind hospitality 
at the NEC Research Institute.



\begin{references}
\bibitem{commgapless}  We do not discuss here the minigaps due to
the curvature of the nanotube surface, see Ref. \cite{Kane-Mele}.
\bibitem{commfbs}  This definition is different from conventional one since
it involves the Fermi points rather than the directions of motion of
scatterred electrons.
\bibitem{comminvar}  The transformed matrix elements $V_{p-p}(2\bar{\alpha}%
K)e^{2i\bar{\alpha}p\chi }$ are invariant under $\pi n/3$ rotations in
chiral angle $\chi _{n}=\chi _{0}+\pi n/3$, which correspond to equivalent
representations of a nanotube.
\bibitem{commRG} We have checked this explicitly for (10,10) SWNTs 
at half filling and away from it.
\end{references}
\end{document}